# Improvement of reversible H storage capacity by fine tuning of the composition in the pseudo-binary systems $A_{2-x}La_xNi_7$ ($A$ = Gd, Sm, Y, Mg)

*Junxian ZHANG, Véronique CHARBONNIER[\*], Nicolas MADERN, Judith MONNIER, Michel LATROCHE*

Univ. Paris-Est Créteil, CNRS, ICMPE, UMR7182, F-94320, Thiais, France

**Abstract**

$A_2B_7$ compounds ($A$ = rare earth or Mg, $B$ = transition metal) are widely studied as active materials for negative electrode in Ni-$M$H batteries. By playing on the substitution rate of both $A$ and $B$ elements, it is possible to prepare various compositions. This strategy will help to improve the properties, for example get a higher reversible hydrogen capacity and a well-adapted hydrogen sorption pressure. Indeed, $A$ can be almost all light rare earths (La to Gd), yttrium and alkaline earth metals (Mg or Ca), whereas $B$ can contain various late transition metals (Mn to Ni). To understand the effects of composition on the physicochemical properties of $La_2Ni_7$-based compound, various pseudo-binary systems have been investigated: $Gd_{2-x}La_xNi_7$ ($x$ = 0, 0.6, 1, 1.5), $Sm_{2-x}La_xNi_7$ ($x$ = 0, 0.5, 1, 1.5), $Y_{2-x}La_xNi_7$ ($x$ = 0, 0.4, 0.5, 0.6, 0.8, 1, 1.5, 1.75) and $A_{0.5}La_{1.1}Mg_{0.4}Ni_7$ ($A$ = Sm, Gd and Y). To determine their crystallographic properties, X-ray diffraction analysis was performed followed by Rietveld refinement. Thermodynamic properties regarding reversible hydrogen sorption were investigated at room temperature. Quaternary compounds present drastically improved sorption properties in the frame of practical energy storage applications with reversible capacity equivalent to 400 mAh/g for the compound $La_{1.1}Sm_{0.5}Mg_{0.4}Ni_7$.

[\*] Current affiliation: Energy Process Research Institute, National Institute of Advanced Industrial Science and Technology (AIST), Tsukuba West, 16-1 Onogawa, Tsukuba, Ibaraki 305-8569, Japan.

1 **Introduction**

The consuming of fossil fuels is not only the main cause of air polluting but also the origin of global warming with more and more alarming consequences. These leads to serious consideration of use of alternative energies: solar, wind and see waves. Those energies require energy storage for final consuming. Hydrogen is one candidate either can be used as fuel[1–3] or charge carrier in Ni-MH batteries[4–6].

$AB_5$ compounds ($A$ = rare earth, $B$ = transition metal) are commonly used as negative electrode materials in commercial Ni-$M$H batteries [4,6,7]. To act as a good anode material, those compounds must exhibit specific thermodynamic properties. First, they must be able to absorb reversibly a large quantity of hydrogen through a large equilibrium plateau pressure. Second, this plateau must lay in the practical electrochemical window located in the pressure range between 0.001 and 0.1 MPa at room temperature [5].

The typical gravimetric capacity obtained with those $AB_5$ compounds reaches around 300-350 mAh/g [5,8,9]. To increase this specific capacity, new systems involving new chemistry are being considered. Indeed, $AB_y$ compounds ($2 < y < 5$) are studied worldwide [9–12]. They can be described as stacking structures made of $[A_2B_4]$ and $[AB_5]$ subunits piled along the $c$ crystallographic axis. Their general formula, introduced by Khan [13], can be written as: $y = \frac{5n+4}{n+2}$ (where $n$ is the number of $[AB_5]$ subunits). For $A_2B_7$, $n$ equals 2 and consequently, the basic period can be written as 2·$[AB_5]$ + $[A_2B_4]$. Those phases are polymorphs as they crystallize either in rhombohedral (3$R$) - Gd$_2$Co$_7$-type - or hexagonal (2$H$) symmetry - Ce$_2$Ni$_7$-type .

Binary compounds Gd$_2$Ni$_7$, Y$_2$Ni$_7$, Sm$_2$Ni$_7$ and La$_2$Ni$_7$ were previously studied [14,15],[16] regarding their ability to form hydrides. These binary compounds absorb large amounts of hydrogen but exhibit several plateaus (two for Gd$_2$Ni$_7$, Sm$_2$Ni$_7$ and La$_2$Ni$_7$; three for Y$_2$Ni$_7$), which is not suitable for application. This behavior is mainly caused by the stacked structure and the mismatch between subunits $[A_2B_4]$ and $[AB_5]$. This mismatch is also responsible for the hydrogen induced amorphization (HIA) of these compounds. Since the atomic size of rare earths play an important role on this geometrical parameter, it is interesting to study the effect of different rare earths on the hydrogenation properties.

In the present paper, we study the effect of lanthanum and/or magnesium substitution to the rare earths $A$ = Gd, Sm or Y in $A_2$Ni$_7$ compounds. Atomic radius $r_A$ of La (1.87 Å) is larger than that of gadolinium (1.787 Å), samarium (1.804 Å) and yttrium (1.776 Å). Lanthanum can occupy the $A$ site either in the $[AB_5]$ or in the $[A_2B_4]$ subunits. This substitution generates an adjustment of the mismatch between the two subunits and thus influences the hydrogenation properties. On the other hand, magnesium (1.599 Å) is also large enough to substitute lanthanides or yttrium, but it only occupies the $A$ site in the $[A_2B_4]$ subunits [17,18]. It plays a key role in decreasing the $[A_2B_4]$ subunit volume and stabilizing the structure toward HIA by reducing the atomic radius ratio $r_A/r_B$ [19]. Furthermore, being much lighter, magnesium leads to much higher specific weight capacity.

Herein the series $A_{2-x}$La$_x$Ni$_7$ and La$_{1.1}A_{0.5}$Mg$_{0.4}$Ni$_7$ ($A$ = Gd, Sm or Y) have been synthesized and studied in terms of crystallographic and thermodynamic properties.

## 2 Experimental

The compounds were synthesized from high purity elements (Gd (Alfa Aesar 99.9%), La (Alfa Aesar 99.9%), Mg (Alfa Aesar 99.8%), Ni (Praxair 99.95%), Sm (Alfa Aesar 99.9%) and Y (Santoku 99.9%)). For the compounds without samarium and/or magnesium, the elements were melted stoichiometrically in an induction furnace. To ensure good homogeneity, this step was repeated five times. The obtained ingots were wrapped in a tantalum foil and annealed for one week under argon atmosphere in a silica tube at 1000 °C. The compounds containing samarium were synthesized with a 2-5 wt.% excess of samarium to anticipate metal evaporation. After induction melting, the ingot was crushed into powder (<100 µm). It was pressed into a 2 g-pellet, wrapped in a tantalum foil and annealed under argon atmosphere in a sealed stainless steel crucible for 3 days at 950 °C. For the compounds containing magnesium, a precursor made of rare earths and nickel was first prepared by induction melting. The obtained ingot was crushed into powder (<100 µm). Magnesium powder (<44 µm) was then added with small excess to the mixture. The final powder was pressed into a 2 g-pellet, which was annealed following the same procedure as for samarium-containing compounds.

All samples were characterized by X-ray diffraction (XRD) using a Bruker D8 DAVINCI diffractometer with Cu-K$\alpha$ radiation, in a $2\theta$ range from 20 to 80° with a step size of 0.01°. Experimental data were analyzed by the Rietveld method using the FullProf program [20].

Chemical compositions were determined by Electron Probe Micro-Analysis (EPMA), using a CAMECA SX-100. Prior to analysis, the samples were polished using 3-µm diamond abrasive paste and ethanol solvent on a woven tape.

Pressure composition isotherms (*PCI*) were measured at 25 °C using the Sieverts' method. All compounds were crushed into a 100 µm-powder, introduced into a gauged sample holder and evacuated under dynamic primary vacuum. Five activation cycles were performed for the Mg-containing samples by successive exposure to 2.5 MPa of hydrogen pressure followed by desorption under dynamic primary vacuum at 150 °C.

## 3 Results
### 3.1 Phase determination and structural properties
#### 3.1.1 Gd-based system

$Gd_{2-x}La_xNi_7$ ($x$ = 0, 0.6, 1 and 1.5) were characterized by means of XRD and EPMA. Resulting diffraction patterns are presented in Figure 12. Cell parameters and phase amounts determined from Rietveld refinement (Fig. SI-1) and composition derived from EPMA are gathered in Table 1. Even though the Ni content is slightly lower than expected, the final compositions are very close to the nominal ones. Similar results have been observed for Sm and Y-based systems (see below). The four alloys are all $A_2B_7$-type phases. The binary $Gd_2Ni_7$ crystallizes in a mixture of 2*H* (37%) and 3*R* phase (63%). The hexagonal 2*H* structure content increases to 54% for $Gd_{1.4}La_{0.6}Ni_7$ alloy, whereas $GdLaNi_7$ and $Gd_{0.5}La_{1.5}Ni_7$ are single phase with 2*H* structure (100%). Both cell parameters *a* and *c* increase with La content.

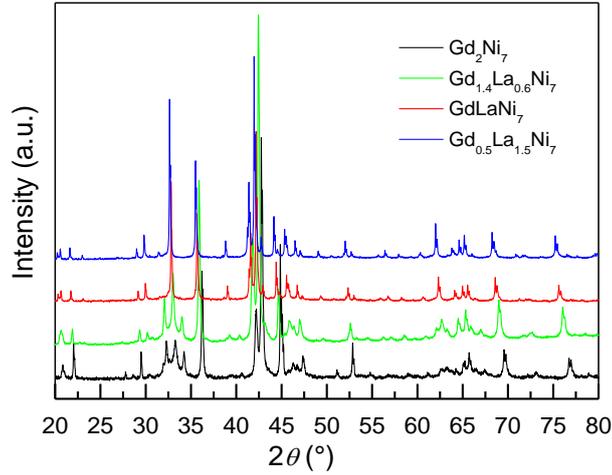

Figure 1: XRD patterns for the Gd$_{2-x}$La$_x$Ni$_7$ system (from bottom up $x$ = 0, 0.6, 1, 1.5).

Table 1: Crystallographic parameters and chemical analysis for the Gd-based system.

| Compound | EPMA Composition ±0.02 | *Rhombohedral - 3R* | | | | *Hexagonal - 2H* | | | |
|---|---|---|---|---|---|---|---|---|---|
| | | $a$ (Å) | $c$ (Å) | $V$ (Å$^3$) | Wt.% | $a$ (Å) | $c$ (Å) | $V$ (Å$^3$) | Wt.% |
| **Gd$_2$Ni$_7$** | Gd$_2$Ni$_{6.93}$ | 4.9608(4) | 36.357(5) | 774.86(7) | 63(3) | 4.9621(8) | 24.230(3) | 516.67 (6) | 37(3) |
| **Gd$_{1.4}$La$_{0.6}$Ni$_7$** | Gd$_{1.4}$La$_{0.6}$Ni$_{6.97}$ | 5.0006(4) | 36.504(4) | 790.5(1) | 46(3) | 5.0015(4) | 24.343(3) | 527.4(1) | 54(3) |
| **GdLaNi$_7$** | Gd$_{0.99}$La$_{1.01}$Ni$_{6.91}$ | | | | | 5.0225(2) | 24.423(1) | 533.55(4) | 100 |
| **Gd$_{0.5}$La$_{1.5}$Ni$_7$** | Gd$_{0.48}$La$_{1.52}$Ni$_{6.90}$ | | | | | 5.0446(1) | 24.558(1) | 541.23(2) | 100 |

### 3.1.2 Sm-based system

Figure 13 shows the XRD patterns for Sm$_{2-x}$La$_x$Ni$_7$ ($x$ = 0, 0.5, 1, 1.5). Phase percentages and cell parameters derived from Rietveld refinement (Fig. SI-2) are summarized in Table 2, as well as EPMA results. As with the Gd, the samples are $A_2B_7$-type phases crystallizing in the two polymorphs 2$H$ and 3$R$. The relative amount of each polymorphic structure does not show clear dependence with the La content. Diffraction peaks in the angular range 30°-35° overlapped, which is typical of stacking faults in $A_2B_7$ compounds. Therefore, the amount of rhombohedral structure determined by Rietveld analysis is slightly over-estimated, especially for $x$ = 1.5, where the two peaks (1 0 10) and (0 1 11) are heavily overlapped and hardly noticeable. Lattice parameters increase with La content for both hexagonal and rhombohedral structures.

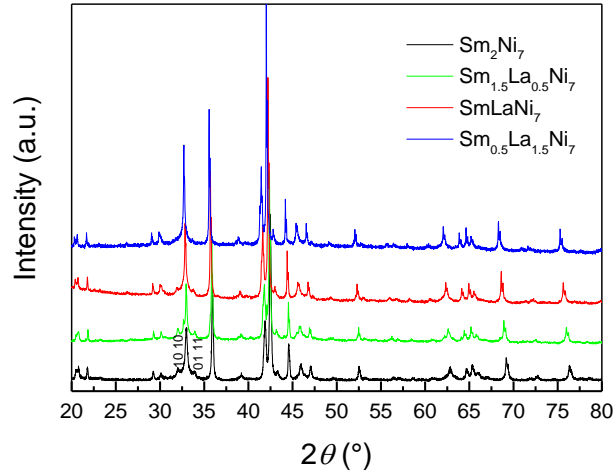

Figure 2: XRD patterns for the system $Sm_{2-x}La_xNi_7$ (from bottom up $x$=0, 0.5, 1, 1.5). The two *hkl* peaks (1 0 10) and (0 1 11) heavily overlapped between 30-35° are shown for $x$=0.

Table 2: Crystallographic parameters and chemical analysis for the Sm-based system.

| Compound | EPMA Composition ±0.02 | Rhombohedral - 3R | | | | Hexagonal - 2H | | | |
|---|---|---|---|---|---|---|---|---|---|
| | | $a$ (Å) | $c$ (Å) | $V$ (Å$^3$) | Wt.% | $a$ (Å) | $c$ (Å) | $V$ (Å$^3$) | Wt.% |
| $Sm_2Ni_7$ | $Sm_2Ni_{6.91}$ | 4.9801(2) | 36.487(2) | 783.70(6) | 24(3) | 4.9800(2) | 24.3290(1) | 522.51(5) | 76(3) |
| $Sm_{1.5}La_{0.5}Ni_7$ | $Sm_{1.52}La_{0.48}Ni_{6.96}$ | 5.0112(1) | 36.59(1) | 795.7(2) | 37(3) | 5.0052(3) | 24.393(2) | 529.23(6) | 63(3) |
| $SmLaNi_7$ | $Sm_{0.97}La_{1.03}Ni_{6.93}$ | 5.0246(1) | 36.696(1) | 802.33(1) | 21(3) | 5.0249(2) | 24.464(2) | 534.94(5) | 79(3) |
| $Sm_{0.5}La_{1.5}Ni_7$ | $Sm_{0.51}La_{1.41}Ni_{6.87}$ | 5.0457(1) | 36.865(1) | 812.83(5) | 15(3) | 5.0451(1) | 24.573(1) | 541.74(3) | 85(3) |

### 3.1.3 Y-based system

A selection of four XRD patterns ($x$ = 0, 0.6, 1 and 1.5) are shown in Figure 14 for the $Y_{2-x}La_xNi_7$ system (all XRD patterns for $x$ = 0, 0.4, 0.5, 0.6, 0.8, 1, 1.5, 1.75 and 2 are given in Fig. SI-3). Table 3 gives the results from Rietveld refinements (Fig. SI-4) and EPMA analysis. $Y_2Ni_7$ crystalizes in pure $Gd_2Co_7$-type phase. Up to $x$ = 0.8, $Y_{2-x}La_xNi_7$ crystallize in a mixture of 3$R$ and 2$H$ phases. For higher La content ($x$ =1, 1.5, 1.75, 2), the compounds crystallize in pure 2$H$ structure. As with the Gd- and Sm-systems, cell parameters (and cell volume) increase with the La content whatever the structures 3$R$ and 2$H$.

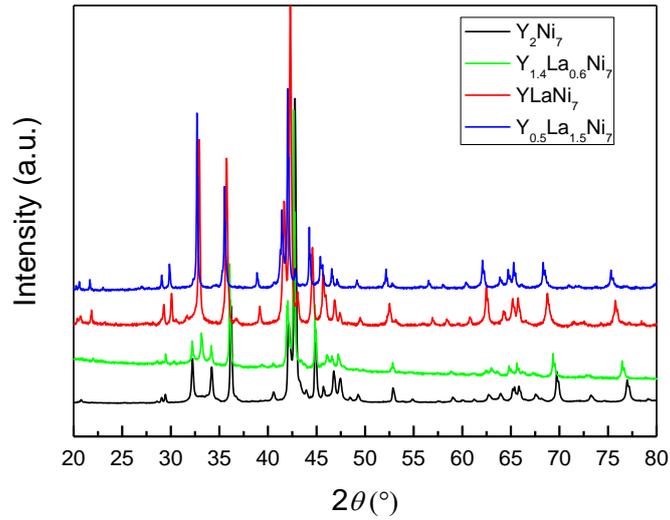

Figure 3: XRD patterns for the system $Y_{2-x}La_xNi_7$ (from bottom up $x$ = 0, 0.6, 1, 1.5).

Table 3: Crystallographic parameters and chemical analysis for the Y-based system.

| Compound | EPMA Composition ±0.02 | *Rhombohedral - 3R* | | | | *Hexagonal - 2H* | | | |
|---|---|---|---|---|---|---|---|---|---|
| | | $a$ (Å) | $c$ (Å) | $V$ (Å$^3$) | Wt.% | $a$ (Å) | $c$ (Å) | $V$ (Å$^3$) | Wt.% |
| **Y$_2$Ni$_7$** | Y$_2$Ni$_{6.92}$ | 4.9465(3) | 36.260(2) | 768.19(7) | 100 | | | | |
| **Y$_{1.6}$La$_{0.4}$Ni$_7$** | Y$_{1.59}$La$_{0.41}$Ni$_{6.83}$ | 4.9800(1) | 36.335(1) | 781.06(3) | 51(3) | 4.9805(2) | 24.244(2) | 520.81(5) | 49(3) |
| **Y$_{1.5}$La$_{0.5}$Ni$_7$** | Y$_{1.49}$La$_{0.51}$Ni$_{6.82}$ | 4.9880(2) | 36.399(2) | 784.26(7) | 56(3) | 4.9904(5) | 24.275(4) | 523.6(1) | 44(3) |
| **Y$_{1.4}$La$_{0.6}$Ni$_7$** | Y$_{1.39}$La$_{0.61}$Ni$_{6.91}$ | 4.9929(2) | 36.412(2) | 786.11(7) | 74(3) | 4.9935(5) | 24.272(3) | 524.1(1) | 26(3) |
| **Y$_{1.2}$La$_{0.8}$Ni$_7$** | Y$_{1.18}$La$_{0.82}$Ni$_{7.01}$ | 5.0072(1) | 36.488(1) | 792.27(1) | 55(3) | 5.0062(1) | 24.317(1) | 527.18(1) | 45(3) |
| **YLaNi$_7$** | Y$_{1.01}$La$_{0.99}$Ni$_{6.86}$ | | | | | 5.0162(3) | 24.3601(2) | 530.82(6) | 100 |
| **Y$_{0.5}$La$_{1.5}$Ni$_7$** | Y$_{0.45}$La$_{1.55}$Ni$_{6.94}$ | | | | | 5.0405(2) | 24.509(1) | 539.27(4) | 100 |
| **Y$_{0.25}$La$_{1.75}$Ni$_7$** | Y$_{0.23}$La$_{1.77}$Ni$_{6.80}$ | | | | | 5.0540(1) | 24.630(1) | 544.83(1) | 100 |
| **La$_2$Ni$_7$** | La$_2$Ni$_{6.86}$ | | | | | 5.0627(4) | 24.714(1) | 548.57(2) | 100 |

### 3.1.4   Mg-substituted compounds

The XRD patterns for the $A_{0.5}La_{1.1}Mg_{0.4}Ni_7$ system ($A$ = Gd, Sm and Y) are shown in Figure 15. Structural analysis are summarized in Table 4 and Rietveld refinements are shown in Fig. SI-5. For sake of comparison, crystal data for $La_{1.5}Mg_{0.5}Ni_7$ are also listed from literature [11]. For all compounds, both polymorphs are observed for these quaternary systems and their cell parameters decrease following the order Sm > Gd > Y.

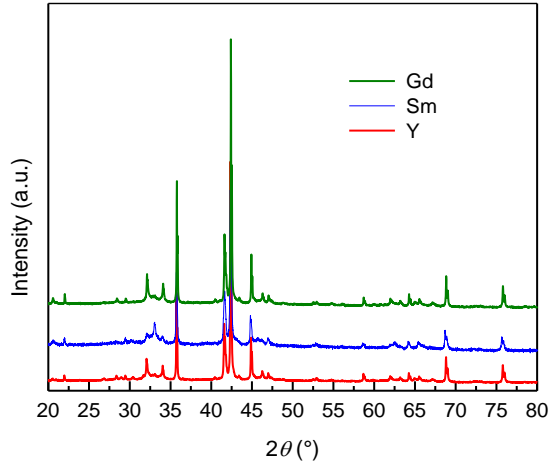

Figure 4: XRD patterns for the $A_{0.5}La_{1.1}Mg_{0.4}Ni_7$ compounds (from bottom up $A$ = Y, Sm and Gd).

Table 4: Crystallographic parameters for the $A_{0.5}La_{1.1}Mg_{0.4}Ni_7$ system ($A$ = Gd, Sm and Y) derived from Rietveld refinements in comparison with $La_{1.5}Mg_{0.5}Ni_7$ obtained from the literature[11]

| Compound | *Rhombohedral - 3R* | | | | *Hexagonal - 2H* | | | |
|---|---|---|---|---|---|---|---|---|
| | $a$ (Å) | $c$ (Å) | $V$ (Å$^3$) | Wt.% | $a$ (Å) | $c$ (Å) | $V$ (Å$^3$) | Wt.% |
| **$La_{1.5}Mg_{0.5}Ni_7$ [11]** | 5.035 | 36.309 | 797.13 | 57 | 5.034 | 24.200 | 531.00 | 43 |
| **$La_{1.1}Gd_{0.5}Mg_{0.4}Ni_7$** | 5.0145(1) | 36.285(2) | 790.16(1) | 74(3) | 5.0148(1) | 24.193(2) | 526.89(1) | 26(3) |
| **$La_{1.1}Sm_{0.5}Mg_{0.4}Ni_7$** | 5.0213(4) | 36.350 (3) | 793.7(1) | 42(3) | 5.0227(3) | 24.252(2) | 529.85(6) | 58(3) |
| **$La_{1.1}Y_{0.5}Mg_{0.4}Ni_7$** | 5.0126(2) | 36.276 (2) | 789.37(6) | 76(3) | 5.0130(2) | 24.182(1) | 526.30(4) | 24(3) |

3.2  Thermodynamic properties
3.2.1     Gd-based system

*P-c* isotherms were recorded for the $Gd_{2-x}La_xNi_7$ system (0≤$x$≤1.5) and are shown in Figure 5 (a). For sake of clarity, all desorption curves are not shown though they were all measured. All compounds exhibit two absorption plateaus and the first plateaus show almost the same hydrogen range between 0.5 and 3.5 H/f.u.. The plateau pressure decreases with La content from 0.08 MPa for $Gd_2Ni_7$ to $3·10^{-4}$ MPa for $Gd_{0.5}La_{1.5}Ni_7$. As for the second plateau, the hydrogen absorption capacity is nearly constant as well. With La content increasing up to $x$=1.5, the PCI shows the same shape as $La_2Ni_7$ with a second plateau at higher pressure 2.2 MPa and a smaller hydrogen content between 5.4 and 9.7 H/f.u.. The hydrogen desorption ranges are between those of $Gd_2Ni_7$ (almost no reversibility) and $La_2Ni_7$ for which the second plateau is partially reversible.

3.2.2     Sm-based system

*PCI* for the $Sm_{2-x}La_xNi_7$ system (0≤$x$≤1.5) are plotted in Figure 5 (b). Like Gd-based compounds, two absorption plateaus are observed. The maximum capacity of the first plateau is 3.5 H/f.u. and does not change significantly with the La content while the plateau pressure decreases significantly from 0.01 MPa for $Sm_2Ni_7$ to $4.6·10^{-4}$ MPa for $Sm_{0.5}La_{1.5}Ni_7$. For the upper (2$^{nd}$) plateau, the capacity increases with $x$La except for $SmLaNi_7$, which shows the lowest capacity. The second absorption plateau pressure decreases slowly with La. For hydrogen desorption, the La substitution improves the desorption ability though it remains incomplete for all compounds.

### 3.2.3 Y-based system

PCI for $Y_{2-x}La_xNi_7$ system ($0 \leq x \leq 2$) are presented in Figure 5 (c) and (d). $Y_2Ni_7$ exhibits three plateaus. The first one at 0.055 MPa which covers a range between 0.4 and 2 H/f.u.; the second plateau is located at 0.5 MPa between 3.1 and 4 H/f.u. and the final plateau at 2.9 MPa between 5 and 8.4 H/f.u. For $0.4 \leq x \leq 0.6$, we observe a plateau at 0.02 MPa up to 3.5 H/f.u., followed by a sloppy branch. While a slight decrease of the first plateau is observed, the second and third ones go down drastically making the PCI curves like a sloppy branch. The compound with $x = 0.4$ exhibits a large capacity of 10.9 H/f.u., and the three compounds show good reversibility. The incomplete desorption are mainly due to the Sieverts' apparatus (small $H_2$ aliquots at very low pressure and limited desorption volume). For $x \geq 0.8$, PCI curves show two plateaus (like for $La_2Ni_7$) but at higher pressures. $Y_{1.2}La_{0.8}Ni_7$ and $YLaNi_7$ show shorter second plateaus, up to 10 MPa, the upper limit of our measurement range, with capacity of 8 and 8.7 H/f.u., respectively. For $x=1.5$ and 1.75, the pressure of the second plateau is clearly reduced and the maximum capacity increases to 10.2 H/f.u.. For the desorption, only one part of the second plateau is observed.

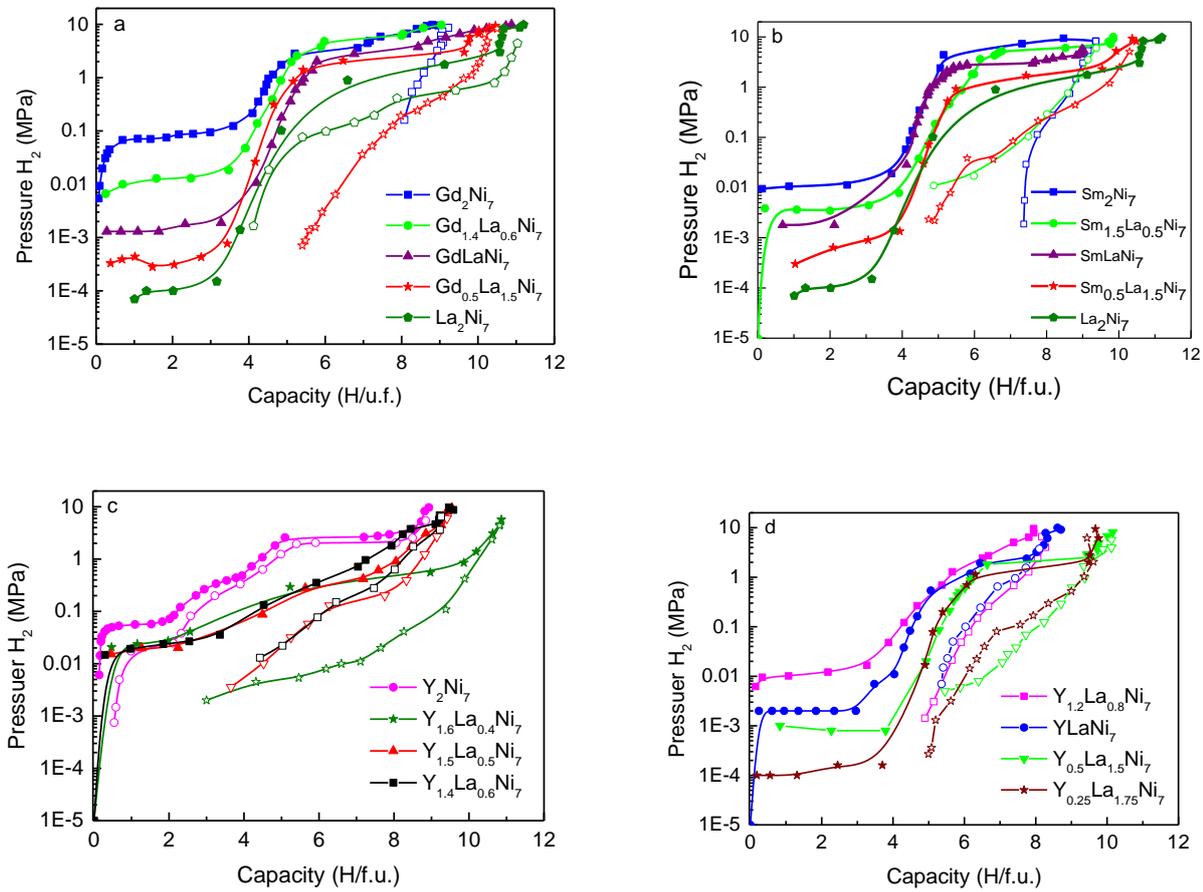

Figure 5: PCI for the system $A_{2-x}La_xNi_7$ with $A$ = Gd (a), Sm (b) or Y (c,d) measured at 25°C, full symbol stand for absorption, open symbols stand for desorption.

### 3.2.4 Mg-substituted compounds

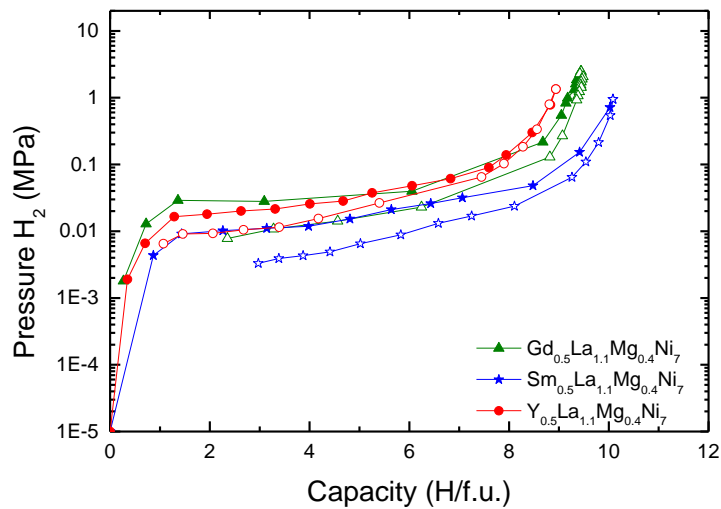

Figure 6: *PCI* for the system $A_{0.5}La_{1.1}Mg_{0.4}Ni_7$ ($A$ = Sm, Gd or Y) measured at 25°C, full symbols for absorption, open symbols for desorption.

PCI curves measured at 25°C for $A_{0.5}La_{1.1}Mg_{0.5}Ni_7$ ($A$ = Sm, Gd or Y) are presented in Figure 17. Outstandingly, all Mg-substituted compounds exhibit only one flat single plateau, with large capacity up to 9~10 H/f.u., in the pressure range between 0.01 and 0.1 MPa and exhibit excellent reversibility.

## 4 Discussion
### 4.1 Structural properties of the pseudo-binary systems $A_{2-x}La_xNi_7$

As expected, the $A$ element (Gd, Sm and Y) can be substituted by La in the whole range for $A_2B_7$ phases, although the abundance of each polymorph changes with composition (see below). On the contrary, for Mg substitution, magnesium can only occupy half of the $A$ sites in the $[A_2B_4]$ subunits, so the maximum Mg content equals to 0.5 for $A_{2-x}Mg_xNi_7$ [18].

EPMA analysis show a slight sub-stoichiometry for Ni in all pseudo-binary $A$-La-Ni systems. Similar results have already been reported in the La(Nd)-Mg-Ni system [21–23], and explained by the presence of possible Ni vacancies or anti-site occupations of Ni by Mg. In the present pseudo-binary systems, it is very unlikely to occupy Ni sites by rare earths (Sm, Gd or Y). Therefore, the presence of some Ni vacancies is hypothesized.

All $A_2Ni_7$ pseudo-binary alloys herein studied adopt either the $Ce_2Ni_7$-type $2H$ or the $Gd_2Co_7$-type $3R$ structure. The two structures differ only in the basic periods stacked along the *c* axis. As a general rule, the larger $A$ atoms favor the formation of the $2H$ structure[24], but the thermal and synthesis history influence also sensibly the formation of either the $2H$ or $3R$ structure[25]. Although the synthesis process slightly differs depending on the composition. Figure 7 shows that for the $A_2Ni_7$ binary compounds ($A$ = Y, Gd, Sm or La), the ratio of hexagonal/rhombohedral structure increases with $r_A$, same effect has been reported in the $(NdA)_{1.5}Mg_{0.5}Ni_7$ ($A$=La,Y) system[26] as welle as (La,Y)$_2$Ni$_7$ system [26,27]. As for the pseudo-binary compounds, the content of hexagonal phase increases with $x$La. For $A$ = Y or Gd, when $x \geq 1$, the compounds are single phase with hexagonal structure. Surprisingly, for $A$ = Sm, which is larger than Y and Gd, the critical value is higher than $x$ = 1.5.

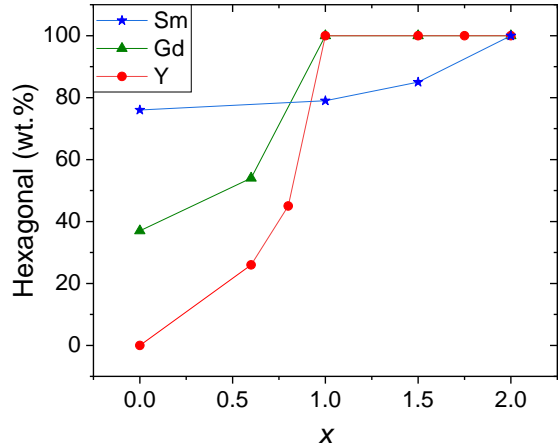

Figure 7: Evolution of the percentage of hexagonal phase in the $A_{2-x}$La$_x$Ni$_7$ systems. The outliers (for $x$ = 0.4 and 0.5) were removed.

For sake of comparison between the hexagonal and rhombohedral structures, the $c$ parameter and the cell volume $V$ have been divided by 6 and 9 for $2H$ and $3R$, respectively, to compare the average parameter $c^*$ and volume $V^*$. Evolutions of the parameters $a$, $c^*$, $V^*$, and $c^*/a$ ratio *versus* $x$La are reported in Figure 19 for the pseudo-binary systems $A_{2-x}$La$_x$Ni$_7$. The cell parameters $a$ ($a_H$ and $a_R$) and $c^*$ increase continuously with La content following a second order polynomial law with opposite curvatures. Besides, a linear increase of the reduced cell volume $V^*$ is observed with $x$La for the three elements $A$. Despite the non-linear behavior of the cell parameters, the opposite effects compensate each other, and the average subunit volume increases linearly with $x$La as expected for a solid solution, following the Vegard's law[28]. The slope factor is logically related to the atomic radius difference between La and the different $A$ atoms following the sequence $\Delta R_{Y-La} > \Delta R_{Gd-La} > \Delta R_{Sm-La}$.

For all $A_{2-x}$La$_x$Ni$_7$ alloys, the $c^*/a$ ratio shows a power law with first a decrease with increasing La content, reaching a minimum for $x$ = 1 and then an increase up to $x$ = 2. This behavior is tightly related to the site occupation of partially substituted La atoms. Refinement of the rare-earth occupancy factors show that at lower $x$, La occupies preferentially the sites $4f_2^H$ ($6c_1^R$) belonging to the [$A$Ni$_5$] subunits. This can be understood based on a geometric effect: La is larger than Y, Sm or Gd, thus it fills the $4f_2^H$ ($6c_1^R$) sites with higher coordination number than the $4f_1^H$ ($6c_2^R$) ones (20 *versus* 16). This result has been confirmed by DFT calculation for the system La$_{2-x}$Y$_x$Ni$_7$, where it was shown that the enthalpies of formation are lower with Y occupying the [$A_2$Ni$_4$] subunit sites and La the [$A$Ni$_5$] subunit one[27]. Replacing Y by La in the [$A$Ni$_5$] slabs leads to an expansion of this slab parallel to the $ab$ plan[15,29]. By epitaxy, this also induces the stretching of the [$A_2$Ni$_4$] in the $ab$ plan, but no effect along the $c$ axis since no substitution arises in this slab. Such anisotropic behavior leads then to a larger cell parameter increase for $a$ than for $c$. This behavior is observed up to the composition La$A$Ni$_7$ ($x$=1), for which La replaced all $A$ atoms in [$A$Ni$_5$]. The stacking structure can then be written as 2·[LaNi$_5$]+[$A_2$Ni$_4$]. Above this composition, La starts to substitute the $A$ sites within the [$A_2$Ni$_4$] slab. The cell parameter $a$ increases slightly, because it has already been expanded within the [$A$Ni$_5$] slab, whereas *c* increases rapidly due to the volume expansion of the [$A_2Ni_4$] slab by La substitution. These hypotheses are confirmed by the evolution of the $c^*/a$ ratio (Figure 19c) that shows a quicker expansion of *a* vs $c^*$ for $0<x<1$ and *vice-versa* for $1<x<2$.

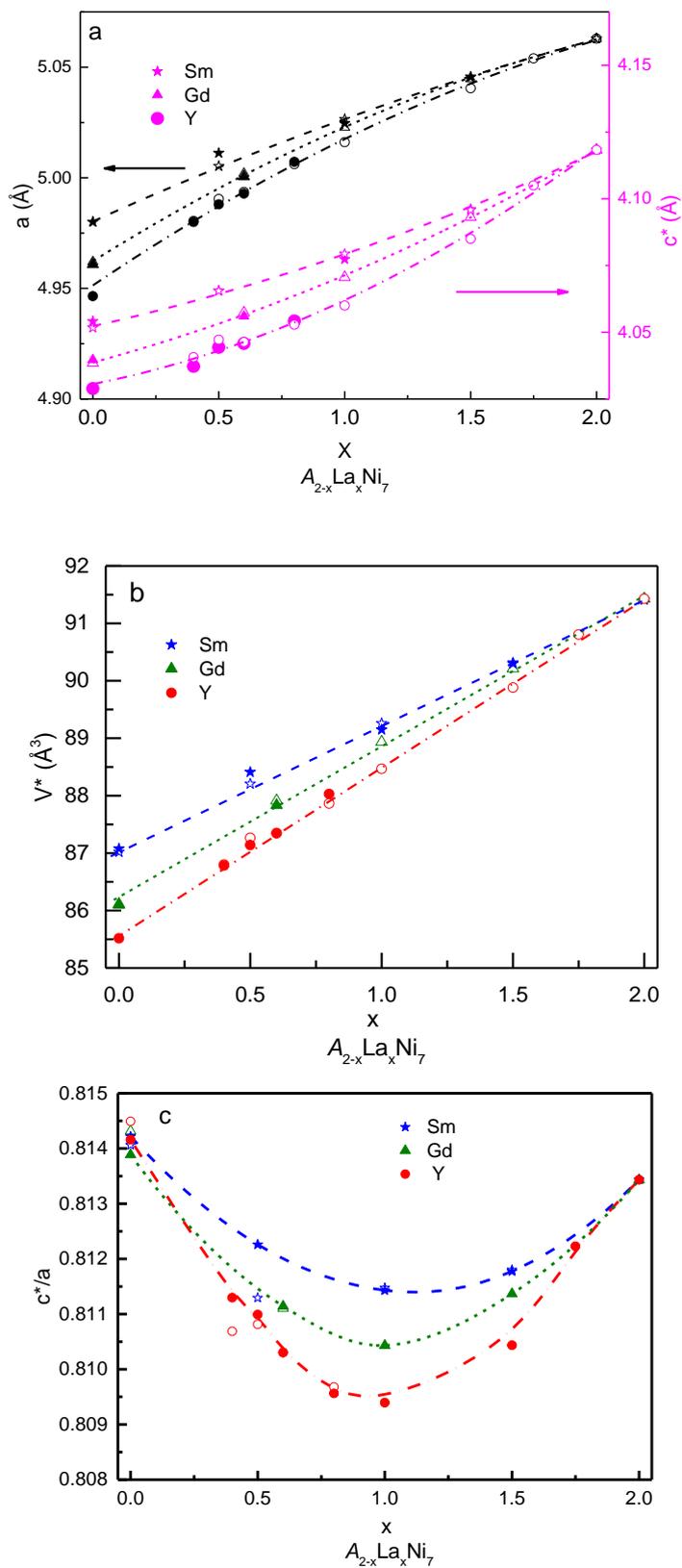

Figure 8 : Evolution of cell parameters *a* and *c\** (a), cell volume *V\** (b) and *c\*/a* ratio versus *x*La content for the pseudo-binary systems $A_{2-x}La_xNi_7$, full symbols for hexagonal structure, open symbols for rhombohedral structure

## 4.2 Hydrogenation properties of the pseudo-binary systems $A_{2-x}La_xNi_7$
### 4.2.1 Comparison of $A_2Ni_7$ binary compounds

PCI for the binary compounds $A_2Ni_7$ (*A* = Y, Gd, Sm or La) are compared in Figure 9. All compounds show two plateaus except $Y_2Ni_7$ with three plateaus. A previous neutron diffraction analysis of $Y_2Ni_7D_z$ has shown that up to the end of the second plateau (i.e. $Y_2Ni_7D_4$), hydrogen fills mainly the available sites in the $[Y_2Ni_4]$ subunit, whereas the third plateau corresponds to the filling of the $[YNi_5]$ one[30]. Anisotropic cell volume expansion has already been observed upon hydrogenation for several $AB_y$ compounds (2<*y*<5) such as $La_2Ni_7D_{6.5}$[31], $CeNi_3D_{2.8}$ [12,32], $Ce_2Ni_7D_{4.7}$ [33], $LaNi_3D_{2.8}$ [34], $CeY_2Ni_9D_{7.7}$ [35] and $Y_2Ni_7D_{2.1}$ [30]. Anisotropic expansion is due to the fact that hydrogen (deuterium) first occupies the $[A_2B_4]$ subunits, for which the *ab* plane is constrained by the surrounding and non-hydrogenated $[AB_5]$ subunits. As those studies pointed out, the individual volume of $[A_2B_4]$ is larger than that of $[AB_5]$, moreover the $[A_2B_4]$ subunits contain 2 *A*-elements instead of one for $[AB_5]$. Thus, $[A_2B_4]$ subunits are richer in *A* and thus more favorable for hydrogen hosting than $[AB_5]$ ones. These differences in geometry and chemistry are the main factors for the multi-plateau behavior of these stacking-structure compounds. Reasonably, we can presume that for binary systems like $Gd_2Ni_7$, $Sm_2Ni_7$ and $La_2Ni_7$, the first plateau corresponds to $[A_2Ni_4]$-H filling whereas the second one is consistent with the filling of $[ANi_5]$.

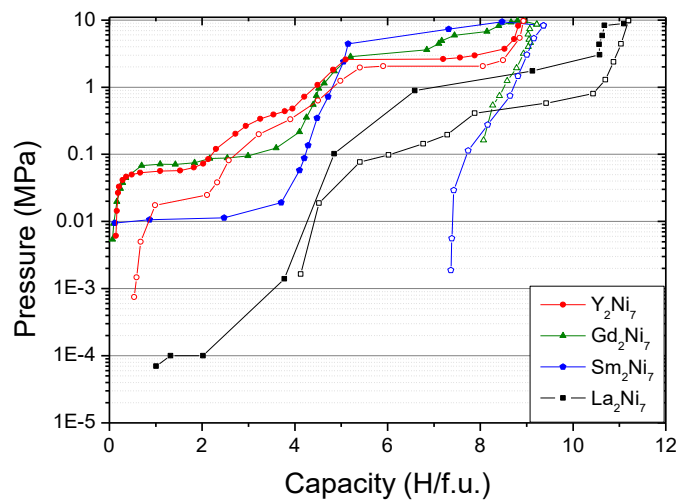

Figure 9: PCI curves for $La_2Ni_7$, $Sm_2Ni_7$, $Gd_2Ni_7$ and $Y_2Ni_7$ measured at 25°C by the Sieverts' method. Filled and hollow symbols correspond to absorption and desorption curves respectively.

Focusing on the first plateau, the equilibrium pressure increases, while the *A* atomic size decreases. $La_2Ni_7$ shows a lower plateau than $Sm_2Ni_7$, which is lower than that of $Gd_2Ni_7$. Indeed, as the rare earths have very similar chemical properties, the $A_2Ni_7H_{\sim 4}$ hydride stability is dominated by geometric parameters. A smaller atom size leads to a smaller cell volume, needing higher energy and consequently higher hydrogen pressure for hydrogenation. However, the first plateau of $Y_2Ni_7$ is slightly below that of $Gd_2Ni_7$, although Y being smaller than Gd. This can be originated by the higher affinity of yttrium for hydrogen, as the formation enthalpy of $YH_3$ is slightly higher than that of $GdH_3$ [36]. Another reason might be related to the difference in electronic configuration due to the 4*f*-electrons of lanthanides compared to *f*-electron free yttrium.

For the second plateau, $La_2Ni_7$ provides the lowest plateau pressure and the highest capacity, while $Sm_2Ni_7$, $Gd_2Ni_7$ and $Y_2Ni_7$ exhibit lower capacities and higher plateau pressures than $La_2Ni_7$. Thus, the hydrogenation properties of those compounds cannot be described by a simple linear combination of the ones of their $AB_2$ and $AB_5$ counterparts. However, assuming that the second plateau corresponds to the filling of the $[ANi_5]$ subunits only, one can expect some relationship between the thermodynamic properties of the second plateau of the $A_2Ni_7$ and $ANi_5$ compounds. Lanthanum is well-known as a good candidate to form hydride in the $ANi_5$ system. At room temperature, $LaNi_5$ shows one flat plateau at 0.17 MPa with a capacity of 6.6 H/f.u. and a very small hysteresis[37]. $GdNi_5$ and $SmNi_5$ show two plateaus, the first one with a capacity of 4 H/f.u. followed by a second one of 2H/f.u. at much higher pressure (above 20 MPa at 25°C)[38,39]. Moreover, the hysteresis between absorption and desorption is quite high for these two compounds. These differences between $LaNi_5$ and both $GdNi_5$ and $SmNi_5$ can explain the large capacity and lower plateau pressure of $La_2Ni_7$ compared to the smaller capacities and higher plateau pressures of $Gd_2Ni_7$ and $Sm_2Ni_7$.

The *PCI* of $Gd_2Ni_7$ and $La_2Ni_7$ have been previously reported[16,40]. For $La_2Ni_7$, our results are in good agreement with the literature. $La_2Ni_7$ transforms from hexagonal to orthorhombic then to monoclinic structure, corresponding to the filling of hydrogen in $[La_2Ni_4]$ and $[LaNi_5]$ subunits, respectively. While the first transformation is irreversible (hydrogen is trapped in $[La_2Ni_4]$ subunits), the second plateau, where hydrogen fills $[LaNi_5]$ subunits, can absorb and desorb hydrogen reversibly. For $Gd_2Ni_7$, a total reversibility at low temperature (258 K) was reported for a sample with a quasi-single hexagonal structure [40]. In our work, we attempted to measure the *PCI* desorption at lower temperature without success due to the sluggish kinetics. As for $Sm_2Ni_7$, it behaves similarly to $Gd_2Ni_7$ showing poor reversibility.

$Y_2Ni_7$ exhibits excellent hydrogen uptake/release reversibility, which is an exception among all $A_2Ni_7$ binary compounds (*A* = Y or lanthanide). Fang *et al.* [41] found that the dehydrogenation properties of $AB_3$ compounds are closely related to their subunit volume. They found that whatever the composition, a $[AB_5]$ subunit volume below 88.3 Å$^3$ and a $[A_2B_4]$ one below 89.2 Å$^3$ are the prerequisite conditions for the reversible hydrogen desorption. For $Y_2Ni_7$, the $[A_2B_4]$ and $[AB_5]$ volumes are 86.1 and 85.1 Å$^3$ respectively, fulfilling the prerequisite conditions for hydrogen desorption according to Fang *et al.*.$A_2B_7$ are compounds with stacking structure, differing from $AB_3$ only by the number of $[AB_5]$ subunits (2 instead 1 for $AB_3$). Their criterion could thus be valid for $A_2B_7$ system too. However, $Y_2Ni_7$ is not the only compound fulfilling these conditions. $Gd_2Ni_7$ exhibits similar crystallographic geometry with $[A_2B_4]$ and $[AB_5]$ volumes of 86.1 and 86.1 Å$^3$, respectively (Figure 11). Under our experimental conditions, reversibility for $Gd_2Ni_7$ was not observed, it might be related to kinetic issues rather than thermodynamic ones.

4.2.2     Effect of the substitution of Sm, Gd or Y by La

Substitution of *A* by lanthanum leads to a decrease of the first plateau pressure. Since La is larger than Gd, Sm and Y, the substitution increases the cell volume and consequently decreases the plateau pressure. As pointed out by Mendelsohn *et al.*, $\ln(P_{eq})$ is directly proportional to the inverse of the cell volume (with $P_{eq}$ the equilibrium plateau pressure) [42]. Equilibrium pressures measured for the first plateau, attributed to the filling of the $[A_2Ni_4]$ subunit, are plotted against the average subunit cell volumes in Figure 21. The $\ln(P_{eq})$ shows a very good linear correlation with the average subunit volume $V^*$. It confirms that the geometrical rule is also valid for the first plateau of those compounds with multi-plateau behavior.

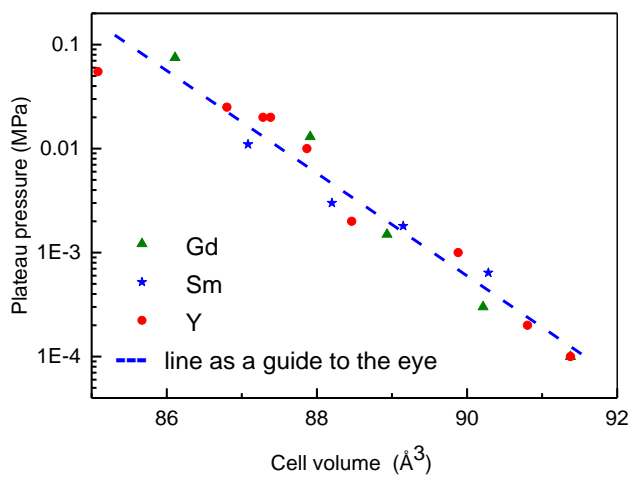

Figure 10: Ln($P_{eq}$) measured for the first plateau as a function of the average subunit volume $V^*$ for $A_{2-x}La_xNi_7$ ($A$ = Gd, Sm, Y).

For the second plateau, the behavior needs more detailed inspection. Surprisingly, for $Gd_2Ni_7$ (Figure 5. a), substitution of Gd by La does not affect significantly the second plateau up to the composition $GdLaNi_7$, which corresponds to the stacking scheme $2\cdot[LaNi_5]+[Gd_2Ni_4]$, *i.e.* substitution in [$AB_5$] unit only. Further increase in La content ($Gd_{0.5}La_{1.5}Ni_7$) leads to a plateau stabilization between those of $La_2Ni_7$ and $Gd_2Ni_7$. In the case of $Sm_2Ni_7$ (Figure 5.b), substitution of Sm by La causes a continuous decrease of the plateau pressure with $x$.

Finally, for $Y_2Ni_7$, La substitution for Y leads to more complex behavior for the second plateau pressure. Low La substitution for Y results in a drastic diminution of the plateau pressure and an augmentation of the capacity. To understand this, the subunit volumes of [$A_2B_4$] and [$AB_5$] are plotted in Fig. SI-6 for the $Y_{2-x}La_xNi_7$ compounds as a function of $x$La. First, when Y is substituted by small amount of La, the [$AB_5$] volume increases rather sharply, which can be explained by the fact that La occupies preferentially the $A$ site in [$ANi_5$] subunit, leading to an expansion of the *ab* plane. Thus, it induces an increase of the [$A_2B_4$] subunit volume, and therefore explain the diminution of the plateau pressure and capacity increase. Up to $x = 0.8$, the subunit volumes of [$A_2B_4$] and [$AB_5$] are still smaller than the reported critical volume (figure 10), which may be the reason for the sloppy PCI curve and the better reversibility. For $x>1$, La starts to occupy the [$A_2B_4$] units, the compounds behave like $La_2Ni_7$. The difference between the volumes of [$A_2B_4$] and [$AB_5$] increases sharply and stands out of the critical limit for reversibility. Those compounds have smaller cell volumes, thus higher plateau pressures compared to $La_2Ni_7$, but similar poor reversibility (Figure 10). However, for fundamental understanding, neutron diffraction of those compounds at different hydrogen contents should be interesting to localize H atoms in the structure.

### 4.2.3 Effect of the substitution of Sm, Gd or Y by La and Mg

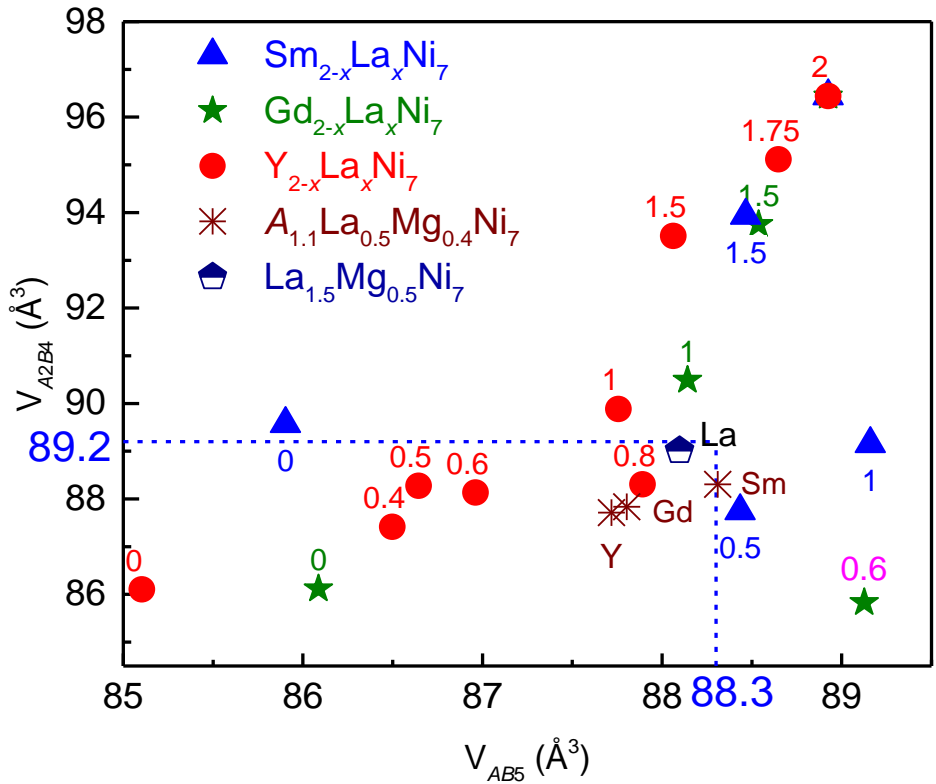

Figure 11: The [$AB_5$] and [$A_2B_4$] subunit volumes for the studied compounds together with the "reversibility limits" giving by Fang *et al.* [41].

PCI of $La_{1.1}A_{0.5}Mg_{0.4}Ni_7$ ($A$ = Sm, Gd or Y) are compared in Figure 17. All Mg-substituted compounds show flat plateaus with good reversibility and low hysteresis. As mentioned before, the mismatch between the [$AB_5$] and [$A_2B_4$] subunits and the too large ratio $r_A/r_B$ are the main causes for the HIA and irreversibility for stacked structure compounds[19,34,43–45]. These authors state that HIA can be suppressed by reducing the subunit volume difference below 5.3 Å$^3$ [46,47]. Recently, Fang *et. al.* pointed out that this criterion is not enough for hydrogen release. They claim that the subunits [$AB_5$] and [$A_2B_4$] should not be larger than 88.3 and 89.2 Å$^3$, respectively. Table 5 summarizes the average subunit and cell volumes of [$AB_5$] and [$A_2B_4$] for the Mg substituted compounds studied in this work and $La_{1.5}Mg_{0.5}Ni_7$, one of the most studied composition with promising hydrogen storage properties[11]. One can see that all these compounds exhibit [$AB_5$] and [$A_2B_4$] subunit volumes very close to each other and fulfill the critical values of 88.3 and 89.2 Å$^3$ respectively (Figure 11), which well explains their outstanding hydrogenation properties. From the point of view of application, these compounds can sorb reversibly large amount of hydrogen at appropriate pressure, ideally flat plateaus in the pressure range 10$^{-3}$ to 0.1 MPa.

Table 5: Subunits ([$AB_5$] and [$A_2B_4$]) volumes, average subunit volume $V^*$ and $c^*/a$ ratio for the Mg-substituted compounds, compared to the values for La$_{1.5}$Mg$_{0.5}$Ni$_7$ [11].

| Compounds | $V_{AB5}$ (Å$^3$) | $V_{A2B4}$ (Å$^3$) | $V^*$ (Å$^3$) | $c^*/a$ | Ref. |
|---|---|---|---|---|---|
| La$_{1.5}$Mg$_{0.5}$Ni$_7$ | 88.09 | 89.01 | 88.40 | 0.804 | Zhang [11] |
| La$_{1.1}$Sm$_{0.4}$Mg$_{0.4}$Ni$_7$ | 88.31 | 88.30 | 88.31(2) | 0.804 | this work |
| La$_{1.1}$Gd$_{0.4}$Mg$_{0.4}$Ni$_7$ | 88.20 | 87.84 | 87.81(1) | 0.804 | this work |
| La$_{1.1}$Y$_{0.4}$Mg$_{0.4}$Ni$_7$ | 87.72 | 87.71 | 87.71(2) | 0.804 | this work |

For the (La,Mg)$_2$Ni$_7$ system, it was reported that Mg substitutes La atoms in the [$A_2$Ni$_4$] units only, up to half of the La sites thus reaching the composition [LaMgNi$_4$] [18,31]. Such effect is generally observed in stacking structures as other authors showed similar behavior with $n$ = 1, 2, 3 (*i.e.* $AB_3$, $A_2B_7$ and $A_5B_{19}$) even with other rare earths [48,49]. The specificity of Mg is therefore not only related to its small size (1.599 Å) but also its localization, as Mg replaces only rare earth located in the [$A_2B_4$] units, thus involving a drastic reduction of the [$A_2B_4$] volume. Consequently, the subunit volumes of [$AB_5$] and [$A_2B_4$] become closer. It is worth to note that, i) Mg has a smaller atomic radius than rare earths, ii) hydrogen affinity for Mg is lower than for the lanthanides and yttrium. Therefore, Mg substitution shrinks the [$A_2B_4$] unit and lowers the H affinity, making [$A_2B_4$] units less attractive for hydrogen. Altogether, it makes the hydrogen occupancies comparable for [$A_2B_4$] and [$AB_5$] subunits leading to larger and flatter PCI curves as observed in Figure 17. It is worth to note that such improvement in the hydrogen sorption properties, combined with the molar mass decrease brought by Mg substitution, leads to materials highly suitable for solid-gas or electrochemical energy storage applications.

The plateau pressures of Gd$_{0.5}$La$_{1.1}$Mg$_{0.4}$Ni$_7$ and Y$_{0.5}$La$_{1.1}$Mg$_{0.4}$Ni$_7$ are very close. Indeed, the volumes of their stacking subunits are almost identical (Table 5). The average subunit volume of La$_{1.1}$Sm$_{0.5}$Mg$_{0.4}$Ni$_7$ is however a bit larger ($V^*_{Sm}$ = 88.31(3) Å$^3$), consequently the plateau pressure is significantly lower. From the PCI, limiting the practical capacity in the pressure range 0.001 to 0.1 MPa, these compounds give good electrochemical capacities with 350 mAh/g for Y, 370 mAh/g for Gd and 400 mAh/g for Sm-based compounds. The relative lower capacity for Y is due to the higher plateau pressure. Electrochemical properties for similar alloys have been studied and shown interesting capacity and cycle life [50,51]. Therefore, by further compositional adjustment of this compound, the pressure can be optimized thus leading to very promising properties for practical applications.

## 5 Conclusions

Several Sm, Gd and Y-based compounds with $A_2B_7$ polymorphic structures have been successfully synthesized in this work. Their structural properties have been investigated and their thermodynamic properties regarding H sorption have been determined at 25°C. For all $A_2B_7$ compounds, the maximal capacity of the fully hydrided samples (up to 10 MPa) ranges between 9 and 11 H/f.u. Comparison of the PCI curves raise several points:

- Substitution of $A$ by La leads to a decrease of the first plateau pressure, the larger the $A$ element in $A_2$Ni$_7$ compounds, the lower the first plateau pressure.
- Substitution of $A$ by La influences the second plateau pressure and the total capacity but it needs more detailed inspection by neutron diffraction to accurately localize H atoms.
- Substitution of $A$ by La and Mg leads to a single, large, stable and reversible plateau pressure, particularly suitable for applications.
- Reversibility can be predicted based on [$A_2B_4$] and [$AB_5$] subunits volumes.

**Acknowledgments**

The authors thank E. Leroy for EPMA measurements, A. Bale and M. Warde for some sample synthesis. This work has been supported by the French ANR (*Agence Nationale de la Recherche*) program PROGELEC under the contract MALHYCE ANR-2011-PRGE-006 01. SAFT and CNRS are also acknowledged for the financial and technical support of the PhD thesis of Dr Charbonnier.

**References**

[1]  Y. Zhu, L. Ouyang, H. Zhong, J. Liu, H. Wang, H. Shao, Z. Huang, M. Zhu, Closing the Loop for Hydrogen Storage: Facile Regeneration of NaBH4 from its Hydrolytic Product, Angew. Chem. Int. Ed. 59 (2020) 8623–8629. https://doi.org/10.1002/anie.201915988.

[2]  B. Sakintuna, F. Lamari-Darkrim, M. Hirscher, Metal hydride materials for solid hydrogen storage: A review, Int. J. Hydrog. Energy. 32 (2007) 1121–1140. https://doi.org/10.1016/j.ijhydene.2006.11.022.

[3]  S. Orimo, Y. Nakamori, J.R. Eliseo, A. Züttel, C.M. Jensen, Complex Hydrides for Hydrogen Storage, Chem. Rev. 107 (2007) 4111–4132. https://doi.org/10.1021/cr0501846.

[4]  N. Kuriyama, T. Sakai, H. Miyamura, H. Tanaka, H. Ishikawa, I. Uehara, Hydrogen storage alloys for nickel/metal-hydride battery, Vacuum. 47 (1996) 889–892. https://doi.org/10.1016/0042-207X(96)00088-7.

[5]  J.-M. Joubert, M. Latroche, A. Percheron-Guégan, Metallic Hydrides II: Materials for Electrochemical Storage, MRS Bull. 27 (2002) 694–698. https://doi.org/10.1557/mrs2002.224.

[6]  Y. Liu, H. Pan, M. Gao, Q. Wang, Advanced hydrogen storage alloys for Ni/MH rechargeable batteries, J. Mater. Chem. 21 (2011) 4743–4755. https://doi.org/10.1039/C0JM01921F.

[7]  T. Sakai, I. Uehara, H. Ishikawa, R&D on metal hydride materials and Ni–MH batteries in Japan, J. Alloys Compd. 293–295 (1999) 762–769. https://doi.org/10.1016/S0925-8388(99)00459-4.

[8]  T. Ozaki, M. Kanemoto, T. Kakeya, Y. Kitano, M. Kuzuhara, M. Watada, S. Tanase, T. Sakai, Stacking structures and electrode performances of rare earth-Mg-Ni-based alloys for advanced nickel-metal hydride battery, J. Alloys Compd. 446–447 (2007) 620–624.

[9]  Y. Liu, Y. Cao, L. Huang, M. Gao, H. Pan, Rare earth–Mg–Ni-based hydrogen storage alloys as negative electrode materials for Ni/MH batteries, J. Alloys Compd. 509 (2011) 675–686. https://doi.org/10.1016/j.jallcom.2010.08.157.

[10] T. Kohno, H. Yoshida, F. Kawashima, T. Inaba, I. Sakai, M. Yamamoto, M. Kanda, Hydrogen storage properties of new ternary system alloys: La2MgNi9, La5Mg2Ni23, La3MgNi14, J. Alloys Compd. 311 (2000) L5–L7.

[11] F.-L. Zhang, Y.-C. Luo, J.-P. Chen, R.-X. Yan, J.-H. Chen, La–Mg–Ni ternary hydrogen storage alloys with Ce2Ni7-type and Gd2Co7-type structure as negative electrodes for Ni/Mh batteries, J. Alloys Compd. 430 (2007) 302–307. https://doi.org/10.1016/j.jallcom.2006.05.010.

[12] V.A. Yartys, A.B. Riabov, R.V. Denys, M. Sato, R.G. Delaplane, Novel intermetallic hydrides, J. Alloys Compd. 408–412 (2006) 273–279. https://doi.org/10.1016/j.jallcom.2005.04.190.

[13] Y. Khan, The crystal structure of R5Co19, Acta Crystallogr. B. 30 (1974) 1533–1537. https://doi.org/10.1107/S0567740874005206.

[14] V. Charbonnier, J. Monnier, J. Zhang, V. Paul-Boncour, S. Joiret, B. Puga, L. Goubault, P. Bernard, M. Latroche, Relationship between H2 sorption properties and aqueous corrosion mechanisms in A2Ni7 hydride forming alloys (A = Y, Gd or Sm), J. Power Sources. 326 (2016) 146–155. https://doi.org/10.1016/j.jpowsour.2016.06.126.

[15] J.-C. Crivello, N. Madern, J. Zhang, J. Monnier, M. Latroche, Experimental and Theoretical Investigations on the Influence of A on the Hydrogen Sorption Properties of ANiy Compounds, A = {Y, Sm, Gd}, J. Phys. Chem. C. (2019). https://doi.org/10.1021/acs.jpcc.9b04600.

[16] K. Iwase, K. Sakaki, Y. Nakamura, E. Akiba, In Situ XRD Study of La2Ni7Hx During Hydrogen Absorption–Desorption, Inorg. Chem. 52 (2013) 10105–10111. https://doi.org/10.1021/ic401419n.

[17] H. Hayakawa, E. Akiba, M. Gotoh, T. Kohno, Crystal Structures of La–Mg–Nix (x=3–4) System Hydrogen Storage Alloys, Mater. Trans. 46 (2005) 1393–1401. https://doi.org/10.2320/matertrans.46.1393.

[18] J.-C. Crivello, J. Zhang, M. Latroche, Structural Stability of ABy Phases in the (La,Mg)-Ni System Obtained by Density Functional Theory Calculations, J. Phys. Chem. C. 115 (2011) 25470–25478.

[19] J. Zhang, G. Zhou, G. Chen, M. Latroche, A. Percheron-Guégan, D. Sun, Relevance of hydrogen storage properties of ANi3 intermetallics (A=La, Ce, Y) to the ANi2 subunits in their crystal structures, Acta Mater. 56 (2008) 5388–5394. https://doi.org/10.1016/j.actamat.2008.07.026.

[20] J. Rodriguez-Carvajal, Fullprof: a program for Rietveld refinement and pattern matching analysis, Phys. B. 192 (1993) 55–69.

[21] J. Zhang, M. Latroche, C. Magen, V. Serin, M.J. Hÿtch, B. Knosp, P. Bernard, Investigation of the Phase Occurrence, H Sorption Properties, and Electrochemical Behavior in the Composition Ranges La0.75–0.80Mg0.30–0.38Ni3.67, J. Phys. Chem. C. 118 (2014) 27808–27814. https://doi.org/10.1021/jp510313a.

[22] V. Serin, J. Zhang, C. Magén, R. Serra, M.J. Hÿtch, L. Lemort, M. Latroche, M.R. Ibarra, B. Knosp, P. Bernard, Identification of the atomic scale structure of the La0.65Nd0.15Mg0.20Ni3.5 alloy synthesized by spark plasma sintering, Intermetallics. 32 (2013) 103–108. https://doi.org/10.1016/j.intermet.2012.09.003.

[23] R. Torres, C. Magen, B. Warot, M. Hytch, J. Zhang, M. Latroche, V. Serin, Quantitative chemical analysis of Ni-MH negative electrode materials by High Angular Dark Field, (2014).

[24] K.H.J. Buschow, A.S. Can Der Doot, The crystal structure of rare earth nickel compounds of the type R2Ni7., J. Common Met. 22 (1970) 419.

[25] A.V. Virkar, A. Raman, Crystal structures of AB3 and A2B7 rare earth-nickel phases, J. Common Met. 18 (1969) 59–66. https://doi.org/10.1016/0022-5088(69)90120-9.

[26] Q. Zhang, B. Zhao, M. Fang, C. Liu, Q. Hu, F. Fang, D. Sun, L. Ouyang, M. Zhu, (Nd1.5Mg0.5)Ni7-Based Compounds: Structural and Hydrogen Storage Properties, Inorg. Chem. 51 (2012) 2976–2983. https://doi.org/10.1021/ic2022962.

[27] V. Paul-Boncour, J.-C. Crivello, N. Madern, J. Zhang, L.V.B. Diop, V. Charbonnier, J. Monnier, M. Latroche, Correlations between stacked structures and weak itinerant magnetic properties of $La_{2-x}Y_xNi_7$ compounds, J. Phys. Condens. Matter. (2020). https://doi.org/10.1088/1361-648X/ab9d4c.

[28] L. Vegard, Z Phys Chem N. F. 5 (1921) 17.

[29] C.E. Lundin, F.E. Lynch, C.B. Magee, A correlation between the interstitial hole sizes in intermetallic compounds and the thermodynamic properties of the hydrides formed from those compounds, J. Common Met. 56 (1977) 19–37. https://doi.org/10.1016/0022-5088(77)90215-6.

[30] V. Charbonnier, J. Zhang, J. Monnier, L. Goubault, P. Bernard, C. Magén, V. Serin, M. Latroche, Structural and Hydrogen Storage Properties of Y2Ni7 Deuterides Studied by Neutron Powder Diffraction, J. Phys. Chem. C. 119 (2015) 12218–12225. https://doi.org/10.1021/acs.jpcc.5b03096.

[31] R.V. Denys, A.B. Riabov, V.A. Yartys, M. Sato, R.G. Delaplane, Mg substitution effect on the hydrogenation behaviour, thermodynamic and structural properties of the La2Ni7–H(D)2 system, J. Solid State Chem. 181 (2008) 812–821. https://doi.org/10.1016/j.jssc.2007.12.041.

[32] V.A. Yartys, O. Isnard, A.B. Riabov, L.G. Akselrud, Unusual effects on hydrogenation: anomalous expansion and volume contraction, J. Alloys Compd. 356–357 (2003) 109–113. https://doi.org/10.1016/S0925-8388(03)00106-3.

[33] R.V. Denys, V.A. Yartys, M. Sato, A.B. Riabov, R.G. Delaplane, Crystal chemistry and thermodynamic properties of anisotropic Ce2Ni7H4.7 hydride, J. Solid State Chem. 180 (2007) 2566–2576. https://doi.org/10.1016/j.jssc.2007.07.002.

[34] R.V. Denys, B. Riabov, V.A. Yartys, R.G. Delaplane, M. Sato, Hydrogen storage properties and structure of La1−xMgx(Ni1−yMny)3 intermetallics and their hydrides, J. Alloys Compd. 446–447 (2007) 166–172. https://doi.org/10.1016/j.jallcom.2006.12.137.

[35] M. Latroche, V. Paul-Boncour, A. Percheron-Guégan, Structural properties of two deuterides LaY2Ni9D12.8 and CeY2Ni9D7.7 determined by neutron powder diffraction and X-ray absorption spectroscopy, J. Solid State Chem. 177 (2004) 2542–2549. https://doi.org/10.1016/j.jssc.2004.03.034.

[36] H. Oesterreicher, Hydrides of intermetallic compounds, Appl. Phys. 24 (1981) 169–186. https://doi.org/10.1007/BF00899753.

[37] C. Lartigue, A. Percheron-Guégan, J.C. Achard, F. Tasset, Thermodynamic and structural properties of LaNi5−xMnx compounds and their related hydrides, J. Common Met. 75 (1980) 23–29. https://doi.org/10.1016/0022-5088(80)90365-3.

[38] H. Senoh, N. Takeichi, H.T. Takeshita, H. Tanaka, T. Kiyobayashi, N. Kuriyama, Hydrogenation Properties of RNi$_5$ (R: Rare Earth) Intermetallic Compounds with Multi Pressure Plateaux, Mater. Trans. 44 (2003) 1663–1666. https://doi.org/10.2320/matertrans.44.1663.

[39] H. Senoh, T. Yonei, H.T. Takeshita, N. Takeichi, H. Tanaka, N. Kuriyama, Appearance of a Novel Pressure Plateau in RNi5-H (R = Rare Earth) Systems, Mater. Trans. 46 (2005) 152–154. https://doi.org/10.2320/matertrans.46.152.

[40] K. Iwase, K. Mori, A. Hoshikawa, T. Ishigaki, Hydrogenation and structural properties of Gd2Ni7 with superlattice structure, Int. J. Hydrog. Energy Optim. Approaches Hydrog. Logist. 37 (2012) 5122–5127.

[41] F. Fang, Z. Chen, D. Wu, H. Liu, C. Dong, Y. Song, D. Sun, Subunit volume control mechanism for dehydrogenation performance of AB3-type superlattice intermetallics, J. Power Sources. 427 (2019) 145–153. https://doi.org/10.1016/j.jpowsour.2019.04.072.

[42] M.H. MENDELSOHN, D.M. GRUEN, A.E. DWIGHT, LaNi5-xAlx is a versatile alloy system for metal hydride applications, Nature. 269 (1977) 45.

[43] S.C. Xie, Z.L. Chen, Y.T. Li, T.Z. Si, D.M. Liu, Q.A. Zhang, Hydrogen absorption–desorption features and degradation mechanism of ErNi3 compound, J. Alloys Compd. 585 (2014) 650–655. https://doi.org/10.1016/j.jallcom.2013.09.153.

[44] K. Iwase, N. Terashita, K. Mori, S. Tsunokake, T. Ishigaki, Crystal structure and cyclic properties of hydrogen absorption-desorption in Pr2MgNi9, Int. J. Hydrog. Energy. 37 (2012) 18095–18100.

[45] J. Chen, H.T. Takeshita, H. Tanaka, N. Kuriyama, T. Sakai, I. Uehara, M. Haruta, Hydriding properties of LaNi3 and CaNi3 and their substitutes with PuNi3-type structure, J. Alloys Compd. 302 (2000) 304–313.

[46] K. Iwase, N. Terashita, K. Mori, S. Tashiro, H. Yokota, T. Suzuki, Effects of Mg substitution on crystal structure and hydrogenation properties of Pr1−xMgxNi3, Int. J. Hydrog. Energy. 39 (2014) 12773–12777. https://doi.org/10.1016/j.ijhydene.2014.06.127.

[47] K. Iwase, K. Mori, A. Hoshikawa, T. Ishigaki, Crystal structure of GdNi3 with superlattice alloy and its hydrogen absorption-desorption property, Int. J. Hydrog. Energy 2011 Asian Bio-Hydrog. Biorefinery Symp. 2011ABBS. 37 (2012) 15170–15174.

[48] A. Férey, F. Cuevas, M. Latroche, B. Knosp, P. Bernard, Elaboration and characterization of magnesium-substituted La5Ni19 hydride forming alloys as active materials for negative electrode in Ni-MH battery, Electrochimica Acta. 54 (2009) 1710–1714. https://doi.org/10.1016/j.electacta.2008.09.069.

[49] L. Lemort, M. Latroche, B. Knosp, P. Bernard, Elaboration and Characterization of New Pseudo-Binary Hydride-Forming Phases Pr1.5Mg0.5Ni7 and Pr3.75Mg1.25Ni19: A Comparison to the Binary Pr2Ni7 and Pr5Ni19 Ones, J. Phys. Chem. C. 115 (2011) 19437–19444. https://doi.org/10.1021/jp2059134.

[50] Z. Cao, L. Ouyang, L. Li, Y. Lu, H. Wang, J. Liu, D. Min, Y. Chen, F. Xiao, T. Sun, R. Tang, M. Zhu, Enhanced discharge capacity and cycling properties in high-samarium, praseodymium/neodymium-free, and low-cobalt A2B7 electrode materials for nickel-metal hydride battery, Int. J. Hydrog. Energy. 40 (2015) 451–455. https://doi.org/10.1016/j.ijhydene.2014.11.016.

[51] C. Tan, L. Ouyang, H. Wang, D. Min, C. Liao, F. Xiao, R. Tang, M. Zhu, Effect of Y substitution on the high rate dischargeability of AB4.6 alloys as an electrode material for nickel metal hydride batteries, J. Alloys Compd. 849 (2020) 156641. https://doi.org/10.1016/j.jallcom.2020.156641.

**Figure caption:**

Figure 12: XRD patterns for the Gd$_{2-x}$La$_x$Ni$_7$ system (from bottom up $x$ = 0, 0.6, 1, 1.5).

Figure 13: XRD patterns for the system Sm$_{2-x}$La$_x$Ni$_7$ (from bottom up $x$=0, 0.5, 1, 1.5). The two *hkl* peaks (1 0 10) and (0 1 11) heavily overlapped between 30-35° are shown for $x$=0.

Figure 14: XRD patterns for the system Y$_{2-x}$La$_x$Ni$_7$ (from bottom up $x$ = 0, 0.6, 1, 1.5).

Figure 15: XRD patterns for the $A_{0.5}$La$_{1.1}$Mg$_{0.4}$Ni$_7$ compounds (from bottom up $A$ = Y, Sm and Gd).

Figure 16: PCI for the system $A_{2-x}$La$_x$Ni$_7$ with $A$ = Gd (a), Sm (b) or Y (c,d) measured at 25°C, full symbol stand for absorption, open symbols stand for desorption.

Figure 17: *PCI* for the system $A_{0.5}$La$_{1.1}$Mg$_{0.4}$Ni$_7$ ($A$ = Sm, Gd or Y) measured at 25°C, full symbols for absorption, open symbols for desorption.

Figure 18: Evolution of the percentage of hexagonal phase in the $A_{2-x}$La$_x$Ni$_7$ systems. The outliers (for $x$ = 0.4 and 0.5) were removed.

Figure 19 : Evolution of cell parameters *a* and *c\** (a), cell volume $V^*$ (b) and $c^*/a$ ratio versus $x$La content for the pseudo-binary systems $A_{2-x}$La$_x$Ni$_7$, full symbols for hexagonal structure, open symbols for rhombohedral structure

Figure 20: PCI curves for La$_2$Ni$_7$, Sm$_2$Ni$_7$, Gd$_2$Ni$_7$ and Y$_2$Ni$_7$ measured at 25°C by the Sieverts' method. Filled and hollow symbols correspond to absorption and desorption curves respectively.

Figure 21: Ln($P_{eq}$) measured for the first plateau as a function of the average subunit volume $V^*$ for $A_{2-x}$La$_x$Ni$_7$ ($A$ = Gd, Sm, Y).

Figure 22: The [$AB_5$] and [$A_2B_4$] subunit volumes for the studied compounds together with the "reversibility limits" giving by Fang *et al.* [41].